\documentstyle[11pt]{article}
\begin{document}

\begin{titlepage}
\begin{flushright}
Freiburg THEP-93/30\\
Heidelberg HD-TVP-94-2\\
gr-qc/9402036
\end{flushright}
\vskip 0.7cm
\begin{center}
\vfill
{\large\bf  ARROW OF TIME IN A RECOLLAPSING QUANTUM UNIVERSE}
\vskip 1cm
{\bf C. Kiefer}
\vskip 0.4cm
 Fakult\"at f\"ur Physik, Universit\"at Freiburg,\\
  Hermann-Herder-Str. 3, D-79104 Freiburg, Germany.

\vskip 0.6cm
{\bf H. D. Zeh}
\vskip 0.4cm
Institut f\"ur Theoretische Physik, Universit\"at Heidelberg,\\
Philosophenweg 19, D-69120 Heidelberg, Germany.\\
\end{center}
\vskip 2cm
\begin{center}
{\bf Abstract}
\end{center}
\begin{quote}
 We show that the Wheeler-DeWitt equation with a
 consistent boundary condition
 is only compatible with an arrow of time that formally reverses
in a recollapsing universe. Consistency of these opposite arrows is
facilitated
 by quantum effects in the region of the classical turning point.
 Since gravitational time dilation diverges at horizons, collapsing
 matter must then start re-expanding ``anticausally"
(controlled by the reversed
arrow) before horizons or singularities can form.
 We also discuss the meaning
  of the time-asymmetric
 expression used in the
 definition of ``consistent histories". We finally emphasize that there
 is no mass inflation nor any information loss paradox
 in this scenario.
 \end{quote}
\vskip 2mm
\begin{center}
{\em Submitted to Physical Review D in February 1994 - still pending}
\end{center}

\end{titlepage}

\section{Introduction}

In conventional statistical physics, the thermodynamical
arrow of time is described by assuming the initial entropy
 to be extremely small compared to its most probable (equilibrium) value,
 while a final state with low entropy is usually rejected as
 ``improbable". Since such a special initial state is, of course,
 equally improbable, this description is not very satisfactory
 as an {\em explanation} of the arrow \cite{1,2} . It is,
 however, entirely consistent with deterministic dynamical laws
 (including the Schr\"odinger equation), since the integration
constants can be chosen by fixing the state arbitrarily at any
time. The sole use of probability arguments would predict an
 uninteresting, thermodynamically symmetric history of the
state with fluctuations around equilibrium for all times.
 Quantum cosmology, on the other hand, does not seem to
admit even the {\em formulation} of such a
 fact-like asymmetry in time \cite{3},
since the classical time parameter disappears in quantum gravity.

In statistical quantum mechanics,
 entropy as the usual measure of time asymmetry is calculated by means
  of the functional
\begin{equation} S[\rho] =-k_B\mbox{Trace}
    \left[\hat{P}\rho\ln(\hat{P}\rho)\right]
    \label{1} \end{equation}
from a density matrix $\rho$ that may correspond to a pure state if
 $\hat{P}=\hat{P}^2$
   is an appropriate trace-preserving projection operator on the space
    of density operators. It represents a concept of {\em relevance}
     or generalized {\em coarse graining} \cite{1}. In order to describe
the time-dependence of the entropy, $\rho$ must here be given in the
Schr\"odinger picture. The increase of entropy is then described
as the transformation of relevant into irrelevant information,
with the corresponding Poincar\'{e} cycles assumed to be much longer
than the age of the Universe.
 Unless this time-dependence of $\rho$
contains also the ``collapse of the wave function" (or,
equivalently, Everett's branching) into {\em definite},
although dynamically indetermined outcomes of quantum measurements,
the formal entropy (1) must include the entropy of lacking
information representing this indeterminism. {\em Physical}
entropy is instead defined as a function of ``given"
classical quantities -- and never as a function of quantities
which would be
represented by any of their superpositions as they may
 arise in measurement-like situations according
to the unitary Schr\"odinger dynamics. If the collapse
represented an asymmetric fundamental dynamical law violating
the deterministic Schr\"odinger equation \cite{4}, it would
 represent an absolute arrow of time and would thus
 be a candidate for the origin of the thermodynamical
arrow. Since there is as yet no experimental hint on such a
fundamental irreversible dynamics and its precise nature,
this possibility will not be considered here.

  As the arrow of time is a cosmic phenomenon, and since the entropy
 of the Universe seems to be dominated by gravity \cite{5}, quantum gravity
  has to be fundamentally included in the description. It will here
   be assumed that the Universe is described by a unified
canonical quantum theory
   that possesses a reparametrization invariant classical
   counterpart such as general relativity. This invariance may
   be interpreted as a kind of Machian principle with respect to time,
   that is, as the absence of any {\em absolute} or preferred
   time parameter \cite{6}. As there are no trajectories in quantum
   theory any more (which could at least be parametrized
    in an arbitrary way), the corresponding Schr\"odinger
   equation for gravity or any other reparametrization invariant
theory can only be of the stationary form
\begin{equation}   H \Psi  =  0 ,    \label{2} \end{equation}
the {\em quantum Hamiltonian constraint} or generalized {\em Wheeler-DeWitt
 equation}. In the absence of any time parameter, this dynamical law does
  not allow one to pose an ``initial" condition of low entropy at {\em any}
   end in time. Although it may be sufficient, and quantum cosmologically
even very plausible, that $\Psi$ depends only on {\em physical}
variables (including all conceivable ``clocks"), the general
nature of the boundary value problem which is required to determine
a solution to the Wheeler-DeWitt equation (2) appears problematic. It may
therefore come as a surprise that, at least in the neighborhood
of Friedmann-type cosmologies, the Wheeler-DeWitt equation (for
constant lapse function $N$) is of the hyperbolic type and thus
defines an intrinsic ``initial" value problem with respect to the
logarithm $\alpha=\ln a$ of the expansion parameter
(or scale factor) $a$ \cite{7}. We shall therefore consider this boundary
value problem as the appropriate way to impose boundary conditions
in quantum cosmology.

  Since the potential that appears in the Wheeler-DeWitt equation (2)
   for these models approaches a simple form in the limit $a \to 0$
  - cf. Eq. (3) below -, it suggests a similar simple structure for its
   solution \cite{8}. While such a simple
   structure can then in fact explain the low
value of entropy at the big bang (including the absence of initial
entanglement and branching), it would do the same for a big crunch
since they both correspond to the region of small $a$ in configuration
space. This would be in agreement with a conclusion reached long ago
by Gold \cite{9} by means of different, and presumably
insufficient, arguments which are based on a classical concept of
time. We emphasize that this is a debate of principle which is
independent of whether or not the Universe will recontract
in reality.

Hawking, on the other hand, has repeatedly claimed \cite{10,11}  --
 following objections by Page \cite{12}
 and Laflamme against his earlier conclusion
\cite{13} in support of Gold - that the
 thermodynamical arrow of time must keep its direction when the Universe
 has reached its assumed maximal extension and starts recontracting.
 His arguments are based on the assumptions of a Wheeler-DeWitt equation
 for closed Friedmann universes together with the Hartle-Hawking
 (``no boundary") boundary condition \cite{14}. This boundary condition
 is based on path integrals (``sums over histories") as a tool
 to calculate the wave function. The same dynamical model will be
 considered here, since it seems indeed to be appropriate for investigating
 these conceptual issues, although we shall avoid using path integrals
 in the definition of the boundary condition for the wave function.
  The no boundary condition
 has to be used with
 caution, since Hawking's conclusions are essentially
 based on semiclassical, or even classical,
  considerations. He has in fact
 explicitly claimed that the no boundary condition
  can {\em only} be used in a semiclassical approximation
(see Ref.~\cite{10} and the discussion following Ref.~\cite{3}).
 It may thus be worth emphasizing
 that approximations (such as WKB) cannot give more reliable results
 than the exact theory -- as plausible as they may appear to our
 classical and time-directed prejudice. If a correct treatment of the
 Wheeler-DeWitt equation (\ref{2}) (which does not present
 unsurmountable technical difficulties in a simple minisuperspace model)
 contradicts the semiclassical results, the latter must be wrong.
 If, furthermore, it is really true that the no boundary proposal can only
 be applied semiclassically, it would then simply be inapplicable
 to {\em quantum} cosmology.

 Hawking, Laflamme and Lyons \cite{11}, for example,
 (HLL for short) state that
  ``the COBE observations indicate that the perturbations which
  {\em  lead to the arrow of time arise at a time}
   during inflation when \ldots Einstein gravity should be a
    good approximation" (our italics).
Quantum mechanically, this description is not
consistent, since the emergence (``arising") of
    quasiclassical properties (including spacetime) from a wave function
 already relies on the most fundamental of all ``irreversible"
 processes (and hence on an arrow of time) -- namely on decoherence.
 Decoherence determines which kind of properties emerge in the form of a
 collapse or branching of the wave function into specific
 ``world components" such as those with definite spacetime geometries
 and, therefore, with definite proper times along all spacetime
 orbits. A symmetric treatment would then give classical time
 an equal opportunity to disappear during the big crunch (if this can
 still be distinguished from the big bang in quantum gravity)
 by means of ``recoherence" \cite{1,15,16,17}.

Our paper is organized as follows. In Section~2 we introduce the
 quantum cosmological model which forms the basis for our discussion.
 We shall then present in detail our arguments which lead us to the
conclusion that the exclusive use of the semiclassical approximation
in \cite{11} is not justified. In Section~3 additional degrees of freedom
are introduced, which is necessary for a discussion of the arrow of time.
We then show that a low entropy condition for the total wave function,
consistently posed
at small scale factor, must lead to a formal reversal of the arrow
of time at the turning point. We also discuss the meaning of
``consistent histories" in this context. Finally, Section~4
focuses on the consequences of this boundary condition for black
holes.

\section{The Quantum Friedmann Universe}

  The Quantum Friedmann Universe is described by a two-dimensional
   minisuperspace (a strongly restricted configuration space) which is
  spanned by the expansion parameter $a$ or its logarithm $\alpha$ and the
   amplitude $\phi$ of a homogeneous scalar field representing matter
   in this model. These variables may then be supplemented by the amplitudes
$x_n$ of all higher multipoles of geometry and matter on the Friedmann
sphere (``perturbations") \cite{18}, which for the conceptual part of the
discussion need not be assumed to be small, and therefore should be able
to describe a realistic quantum universe. The Wheeler-DeWitt equation
in minisuperspace is given by
\begin{equation} \left(\frac{\partial^2}{\partial\alpha^2}- \frac{\partial^2}
{\partial\phi^2}-e^{4\alpha}(1-e^{2\alpha}m^2\phi^2)\right)
\psi(\alpha,\phi)=0. \label{3}  \end{equation}
This is equivalent to Eq. 2.6 of HLL,
 except for a different factor ordering that is irrelevant
 for our discussion.

  Any boundary condition which is imposed to determine a solution to (3)
  clearly has to be understood as a condition to fix the
  (stationary) {\em wave function},
   even if it is technically expressed by means of (parametrizable) Feynman
  {\em paths}. These paths have no a priori {\em physical} meaning
  in quantum theory. Their superposition in a path integral, in particular,
  always represents a wave function and never merely an ensemble of
  paths or of (consistent or inconsistent) ``histories." The central
  point of our argument is that the mathematical boundary value
  problem for the wave function $\psi$ (and its generalization
  to full superspace) does neither ``know" of its intended physical
  interpretation, nor of any external or derivable concept
  of time that could possibly {\em distinguish} initial from final
  boundaries. Physical conclusions should then only be drawn from the
  solution of this equation, which must {\em depend} on its exact
  boundary condition, and may turn out to be {\em approximately}
  compatible with parametrizable trajectories {\em in certain regions}
  of configuration space where concepts of geometrical optics
  are applicable, i.e., where narrow wave packets which form
  local {\em components} of the solution
  propagate without dispersion. (Such wave packets may decohere
  from one another when further degrees of freedom are taken
  into account.)
   Obviously, this seems to be the case in that region
  which represents the present era of our observed Universe, but may
  not necessarily hold everywhere in superspace -- including that part
  which corresponds to our Universe's early past or late future
  (as far as they are defined at all in timeless quantum gravity).
  The concept of trajectories with their own ``time" parameters
  (perhaps to be justified by a saddle point approximation in the
  path integral representation) must, in particular, not be assumed
  to be applicable for $a\to 0$, where the
  cosmological boundary conditions of low entropy are
  to be imposed.

   A conventional way of solving (3) would consist in specifying
   $\psi$ and its derivative $\partial\psi/\partial\alpha$
    on a hypersurface of fixed
    $\alpha=\alpha_0$, and then integrating with respect to
    $\alpha$. This procedure (which in particular is indicated by the
     hyperbolic nature of this equation) could be used to impose
     an ``initial" condition at small
 $a$, for example, but it would then in general {\em not}
 lead to a reflection of the wave from the repulsive curvature potential
 $e^{4\alpha}$ at very large values of $\alpha$.
  One would instead obtain
 an exponentially increasing wave function in the ``classically forbidden"
 region (as was in fact claimed, for example, by Hawking and Wu
 \cite{19}).
A reflection at large $\alpha$, however, seems to be needed if one wants
to describe a recollapsing universe in quantum cosmology.
Although it is far from clear what kind of ``norm" one has to use for
a wave function in quantum cosmology\footnote{An interesting
proposal to use the Klein-Gordon norm in this context has recently been
made in \cite{HW}. In its present formulation, however,
this proposal has only been applied to describe expanding universes.
If a Klein-Gordon current (the same holds for a Schr\"odinger
current) hit a potential barrier without a condition of
exponentially decreasing wave functions being imposed
in the ``forbidden" region,
it would continue undamped into the region inside the barrier.
No ``corrections" would then be required in the allowed
region.},
we regard it as appropriate to proceed in accordance with standard quantum
mechanics and to {\em exclude} exponentially {\em in}creasing
solutions in ``classically forbidden" regions. (Note that we do not
impose the stronger assumption of ${\cal L}^2$ integrability
in $\alpha$.) Although there are no strictly classical forbidden
regions due to
 the indefinite nature of the kinetic energy, without this condition
  one would not at all be able to represent the trajectory of
 a recollapsing universe in quantum mechanical terms, that is,
  by a propagating wave packet (or stationary ``wave tube").
 This boundary condition of excluding exponentially increasing wave functions
 for large $\alpha$
 restricts the freedom of choice for the
 ``initial" conditions of this second order equation (again in terms of
 the variable $\alpha$) to only {\em one} remaining function of
 $\phi$ (and the amplitudes $x_n$ in full superspace). It thereby
 constrains the solution to contain the reflected partial waves
 ``from the beginning" \cite{16,20} --
 precisely as the boundary condition at $r=0$ in conventional scattering
 problems requires outgoing waves to be present at $r\to\infty$.
  Wave mechanically these
 two parts of the wave function must interfere, and cannot,
 in general, be conceptually separated from one another, since there
 is no dependence on an external time. Even decoherence can only
 occur as a process with respect to an ``intrinsic time" $\alpha$.
 In contrast to the stationary treatment of conventional scattering
 problems, which is justified by the construction of time-dependent
 wave packets, the stationary form of the Wheeler-DeWitt equation
 is exact.

  Note that the terms ``initial" and ``final" have here quite different
   meanings with respect to classical and quantum dynamics. With respect to
    the wave equation they refer to increasing
    $\alpha$, while the classical language refers to parameters of
 (possibly returning) orbits, which do not exist in quantum theory.
Since the Wheeler-DeWitt equation is fundamentally stationary and real,
there is no absolute time variable which could give rise to a reference
phase $e^{i\omega t}$ that might enable one to distinguish between
``incoming" and ``outgoing" parts $e^{\pm ik\alpha}$ of the wave function.
The signs of all momenta can then only have relative meaning.

 The ``no boundary" condition cannot easily be transformed into the
  wave mechanical form appropriate for directly solving the
  Wheeler-DeWitt equation. HLL claim that it corresponds
   to a ``boundary condition at {\em one} end of the four geometry",
 that is, to a {\em classical} initial or final condition.
 In minisuperspace this condition is claimed to read
 (see their Eq. (2.8))
\begin{equation} a=0,\; \frac{da}{d\tau}=1,\; \phi=\phi_0,\;
    \frac{d\phi}{d\tau}=0, \label{4} \end{equation}
where $\tau = it$ is an imaginary time parameter. Because of the fourth
 condition, a real time parameter $t$ would, in spite of the generally
 indefinite sign of the kinetic energy of gravity, lead to a sign which is
 incompatible with the potential $V(a,\phi):=-e^{4\alpha}
 +e^{6\alpha}m^2\phi^2 <0$ in the Planck era. Quantum mechanically,
 this is known to require a wave function depending exponentially
 on $\alpha$ if $p_{\phi}^2$ can be neglected in correspondence
 to the fourth condition in (4). While the first condition defines
 the boundary, the third one merely labels the starting points
 of the paths on it. Since all paths are furthermore assumed
 to contribute with equal initial amplitude (at $\alpha\to-\infty$),
 they form a boundary condition $\psi(-\infty,\phi)=constant$,
 which is compatible with the fourth condition in (4).
 Other boundary conditions at $\alpha\to-\infty$ may correspond to
 initially ``superluminal" trajectories with $\vert d\phi/dt\vert
 >\vert d\alpha/dt\vert$, parametrized by real values of $t$.
As mentioned above, there are no absolutely forbidden regions,
but {\em certain} trajectories will be reflected, while others are
bent to superluminal angles. Analogous results are obtained for propagating
wave packets in the WKB approximation {\em if} the exponentially
increasing solutions are excluded.

 Instead of solving the wave function, however, HLL follow
 {\em trajectories} which
  start at $\alpha=-\infty$ and are parametrized by
{\em imaginary} values of time until they reach the border line represented by
$V(a,\phi) = 0.$\footnote{The only consequence of the imaginary time
parameter in this ``forbidden"
 region of configuration space is that the path integral
leads to an exponential WKB wave function (which is here {\em chosen}
to increase with $a$ along these pseudotrajectories
-- in precise analogy to what we require in the
``forbidden" region for large $a$). In the limit
$\hbar\to 0$ the penetration depth for the wave function into this
region would vanish. It therefore appears misleading to interpret
this situation as describing ``Euclidean spacetimes."
When conventional quantum theory is applied to gravity, the wave function
 describes nothing but a stationary
 probability amplitude for three-geometries.}
 They then ``continue" these trajectories by means of a real
 time parameter (corresponding to an oscillating WKB wave function)
 through the whole history of a universe and obtain in this way
 quite different final conditions at the other end of these
 classical paths. Such a classical condition should,
 however, already be part of the wave mechanical ``initial"
 condition which has to fix the wave function completely.
 These final conditions, derived by classical methods,
 are interpreted by HLL as defining ``corrections" to the wave function
 for small $a$, although they must possess the same measure
 as the original component in an appropriate sense, for example in terms
 of the conserved Klein-Gordon current.

 This resulting asymmetry of the
 {\em  trajectories} (albeit not defined with respect to any
 {\em  absolute} sense of time) is the basis of their conclusion that the
  arrow of time must continue beyond the turning point. Although classically
   consistent, it is here a consequence of the asymmetric treatment
of both ends (motivated by the usual interpretation of the path integral
as a propagator in time), and not just of the use of trajectories
-- which especially in the Planck era is a doubtful concept.
A symmetric treatment of paths or trajectories would select those
(much fewer) ones which obey equivalent conditions at both ends
-- although they would not have to be individually symmetric
(as had originally been expected by Hawking \cite{13}). The direction
of calculation cannot be based on the direction of any ``causality"
still to be derived. This is true even for the construction of a WKB
wave function that includes a second sheet describing the reflected
wave.

 If the asymmetric {\em procedure of calculation} did in fact lead to
  ``corrections" to the wave function at small $a$, this would simply
  demonstrate that the original boundary condition for small $a$ is
 incompatible under the assumed wave equation with that required
 at large $a$ for describing the reflection. There would, however, be
 no justification for this ``final" condition affecting only {\em one}
 end of the quasitrajectories, which quantum mechanically have to be
 represented by propagating wave packets. The exact
 Wheeler-DeWitt equation does not even provide conceptual means
 to apply a boundary condition of the form $\psi(-\infty,\phi)
 =constant$ to only one ``sheet" of the wave function, since the
 concept of separate sheets is facilitated only by the nonlinearity
 of the Hamilton-Jacobi equation. Any insistence on semiclassical
 concepts as being essential for the interpretation of the theory
 would demolish its claim as representing {\em quantum}
 gravity. In the Planck era, for example, trajectories are no better
 justified than in a hydrogen atom.

 How may one consistently interpret the path integral used in the
  no boundary proposal? If it were to represent wave propagation according
 to the ``Klein-Gordon dynamics" (3) from some ``initial time"
 $\alpha_0$ to, say, $\alpha_1$, the paths would have to be
  parametrized by $\alpha_0\leq\alpha\leq\alpha_1$. This propagation
  would then be defined regardless of any reflection from the
  repulsive potential at large $\alpha$ (although an arbitrary initial
condition would in general be inconsistent with the corresponding
final condition if the wave function were propagated that far).
Such an interpretation of the path integral is subtle and even
questionable,
since a composition law,
as it must of course be valid for the
wave equation (3), does no longer seem to hold for the
quantum cosmological path integral
 \cite{21}. The path integral quantization has,
therefore, occasionally been suggested to be more general
than the canonical (wave function) quantization \cite{22}.
Such a ``generalized quantum theory" would, however, be
speculative, and transcend the realm of the empirically confirmed
theory. The usual evaluation of path integrals by means of a WKB
approximation, on the other hand, does not even correspond
to the Heisenberg-Schr\"odinger quantization, since it represents
the Bohr-Sommerfeld level of quantum theory.

 One may thus rather interpret the path integral in a first step as
 representing a propagation with respect to a {\em formal} parameter
 $t$ according to
 $\Psi(t) = e^{-iHt}\Psi(0)$, where $H$ is the Wheeler-DeWitt Hamiltonian,
  followed by a projection onto the corresponding zero frequency mode,
  that is, onto a solution of the Wheeler-DeWitt equation
  $H\Psi=0$ by integrating over $t$ from $-\infty$ to $+\infty$
  \cite{23,24}. In minisuperspace this construction would
  read explicitly
\begin{equation} \psi(a,\phi)= \int_{-\infty}^{+\infty}dt\int
    da'd\phi'G(a,\phi;a',\phi';t)\psi^{(0)}(a',\phi'), \label{5}
    \end{equation}
with the formal propagator $G(a,\phi;a',\phi';t)$ which
  would propagate a formal ``initial" wave function
 $\psi^{(0)}$
 that has to be given on the full configuration space
  (including {\em all} values of $a$) --
not only on a boundary. Although this initial function may be
  {\em  chosen artificially} to contain a factor $\delta(a)$,
perhaps multiplied by a constant in $\phi$, such an assumption
would not represent a natural choice of a boundary condition for this formal
``dynamics." Any information about a direction of propagation
in formal time $t$ would be lost by the integration.
The no boundary proposal thus can only yield wave functions
which are given ``at once" on the full configuration space. We also note
that the construction (5) with a $\delta$ function as
mentioned above, if calculated exactly in simple models,
does not allow the construction of narrow wave packets which follow classical
trajectories \cite{24}.

It is clear that classical
  {\em trajectories} in minisuperspace (derived from a boundary
  condition or not) are generically asymmetric. However, they
   would merely represent a
  situation well known from classical mechanics, where solutions
 from a symmetric Lagrangean are also {\em not}
 symmetric, without in general offering
  any thermodynamical insights. The solutions considered by Laflamme
  and Shellard \cite{25} for Kantowski-Sachs universes, for example,
  were chosen to {\em start} at or near a disk-like singularity, and must
  then evolve into a cigar-like one.
In contrast to the recollapsing
 Friedmann minisuperspace trajectory, these initial
and final singularities are distinct in configuration space.
(In the vacuum case, the disk-like singularity would merely
represent a topological one, caused by the chosen foliation of the
upper Kruskal wedge according to the Schwarzschild coordinate $r$.
Classical solutions with matter may similarly ``bounce"
from the disk-shape for earlier times.)

   In special situations, a Klein-Gordon type equation
   (here with variable ``mass term") may nonetheless be consistent
    with the concept of initial conditions for {\em reflected trajectories}
    (which would then allow to impose very different boundary conditions for
     {\em additional} degrees of freedom
     -- see Section~3). This would require that, first,
     geometrical optics is applicable with sufficient precision
along the {\em whole} trajectory, and
 second, that there is a region on a spacelike ``initial" hypersurface
 in minisuperspace to which reflected quasitrajectories, which started
 there, never return. A simple example is a plane timelike potential
 barrier in Minkowski space, hit nonorthogonally by a spatially bounded
 wave packet (for this purpose equivalent to an ensemble of
 trajectories). It may easily be constructed from plane waves
 which fulfil the boundary condition of vanishing at the barrier.
 This would allow completely free initial conditions on a {\em partial}
 ``initial" (or ``source") region, and thereby determine the wave function
 in the resulting disjoint ``final" (or ``image") region on the same
 Cauchy surface (such as $\alpha=constant$). The initial region
 would itself have to be selected by that half of the complete initial
 conditions which remain free after the barrier condition at large
 $\alpha$ has been imposed, while the final region would then be determined
 in this way.

  Even this ad hoc distinction between ``initial" and ``final"
   regions according to this special ensemble of classical solutions
   or ``light rays" fails wave mechanically if the wave packets
   show sufficient dispersion that prevents them from remaining
   disjoint. Precisely this turns out to be the case in the minisuperspace
characterizing closed Friedmann universes, where reflected wave
packets are found to be scattered over their whole configuration
space as a consequence of the specific form of the repulsive
curvature potential \cite{3,20}. This dispersion demonstrates that the
semiclassical approximation for (\ref{3}) {\em cannot} be valid
all around the region of a classically expanding {\em and}
recollapsing trajectory.
 Quantum effects are thus essential
not only in the ``Planck era."

\section{Decoherence and ``Consistent Histories"}

 If the two partial wave packets which formally represent the
  ``expanding" and the ``collapsing" Universe intersect or overlap in
   minisuperspace (as they do
repeatedly in the two-dimensional quantum Friedmann model
    even without any dispersion), they must interfere unless they
    decohere from one another. If the environmental degrees of
freedom that contribute to this decoherence can themselves be
described by a WKB approximation, this decoherence simply means
that the partial wave packets, which are then dispersion-free,
travel in disjoint slices of the complete configuration space,
and hence do not overlap any more. In general, however, decoherence
results from quantum scattering processes which follow an arrow of time
determined by a Sommerfeld radiation condition of negligible
{\em initial} correlations. It would therefore represent circular
reasoning to continue the statistically interpreted collapse of the wave
function along a trajectory describing a reversal of the expansion
of the Universe in order to {\em derive} a continuing thermodynamical
arrow of time. Instead, one has to expect recoherence (derived
from an inverse Sommerfeld condition) to occur there.

 The arrow of time requires statistical considerations,
  and therefore additional degrees of freedom (such as the
   ``environmental" ones which are responsible for decoherence).
 In the Friedmann model, the entropy may be defined by
 means of the functional (1) from the wave functions $\varphi
 (\alpha,\phi;\{ x_n\})$ for the higher multipoles $x_n$ defined by the
 ansatz \cite{18,26}
\begin{equation} \Psi(\alpha,\phi,\{ x_n\})= \mbox{Re}\left(\psi(\alpha,\phi)
    \varphi(\alpha,\phi;\{ x_n\})\right). \label{6} \end{equation}
$\varphi$ is here assumed to depend only slowly on
$\alpha$ and $\phi$, while $\psi$ may, in certain regions, be
 approximated by a WKB solution of the form $e^{iS(\alpha,\phi)}$
with a Hamilton-Jacobi function $S$.
  The ansatz (6) is more general than a product of individually
   real factors \cite{27}. In particular, one may derive from it a
   ``time-dependent Schr\"odinger equation" for the multipoles,
\begin{equation} i\frac{\partial\varphi}{\partial t}:= i\nabla
S\cdot\nabla\varphi
    \approx H_{eff}\varphi, \label{7} \end{equation}
that is approximately valid along the parametrized
 trajectories $\phi(t), \alpha(t)$ in minisuperspace defined by means
  of the gradient $\nabla S$. The orbit parameters $t$
   (or ``WKB times") assume the role of time as a
    ``controller of motion" for this effective dynamics.
     This ``complexification" of the time-dependent Schr\"odinger equation,
which is derived from the {\em real} Wheeler-DeWitt wave function,
represents a strong spontaneous symmetry breaking as it is typically described
by
 means of a nonlinear approximation in quantum theory \cite{28,29}.

In order to be compatible with the conventional quantum mechanical
 description of the observed world, Eq. (7) must describe measurement
  processes according to von Neumann's unitary (``second") dynamics
 in a time direction of growing entropy and entanglement. In addition
to physical entropy, the statistical entropy calculated from
$\varphi(t,\{x_n\})$ by means of (1) must therefore contain the
entropy that measures the missing information about the outcome of all
measurement-like processes which occurred in the respective ``past"
of $t$. The time-dependent Schr\"odinger equation must hence {\em not}
be used to determine the wave function describing the state in our past
by calculating backwards in time, starting only
with a ``branch" wave function
that represents the present state of the ``observed world" (with definite
classical properties). Such a calculation would miss the deterministic
predecessors of those ``non-observed Everett components" that are
physically meaningful and important in forming superpositions
which may define {\em observed} past states. For precisely the same reason,
the Schr\"odinger equation must then also not be used to calculate
the formal ``future" of $\varphi$ beyond the turning point of the cosmic
expansion {\em if} the arrow of time reverses at this point.
On returning quasitrajectories in minisuperspace, one would expect
anticausal and nonunitary contributions to (7) to occur for the
same reason (the boundary condition) which leads to
causality and branching during expansion. In this case the {\em physical}
direction of time is reversed with respect to the formal
parameter of the trajectory. Eq. (7) can neither be always
meaningful nor exact if $\varphi$ is defined by Eq. (6).

If (7) is instead used to {\em define} $\varphi$ along the complete
 trajectory \cite{30}, starting with an ``initial" state of low
 entropy (for example at the same end as used to apply (4)), an
  ever increasing entropy will result for all times smaller than
the corresponding Poincar\'{e} times (which would greatly exceed
any conceivable duration of the Universe). In contrast, solutions of the
Wheeler-DeWitt equation of the form (6) with relative states
$\varphi(\alpha,\phi;\{x_n\})$ of low entropy everywhere for small
$a$ would necessarily describe a reversing arrow along turning
quasi-trajectories. Close to the turning point there would be a region
of indefinite de- or recoherence, that is, a region that cannot
be interpreted in classical terms (similar to the Planck era).

Wave packets constructed from that half of the solutions of
 the Wheeler-DeWitt equation which obey the boundary condition at
  large $a$ must automatically render ``initial" and ``final"
  conditions (in the sense of trajectories) quantum
   dynamically compatible with one another. They cannot, in general,
    form complete quasiclassical histories, though, since they must
    represent {\em superpositions} of many very different classical
    worlds at least on one leg of their histories because of the
quantum scattering that occurs at the turning point \cite{26}.
These superpositions must decohere into very {\em different}
branches on both legs, i.e. there does not
exist any classical connection between different
legs across some ``turning point".

Can such wave packets then at least consistently describe quasiclassical
 worlds during {\em one} of their halfcycles? Low entropy conditions at both
  ends of quasiclassical time would be compatible with the observed
   time asymmetry if we happened to live close to one end, and if the
world were ``informationally opaque" somewhere during its complete
history (for example, by closely approaching thermodynamical
equilibrium in the middle), so that no information could survive the
turning point of the expansion \cite{31}. The second condition has been
questioned to be realistic \cite{32}, since our Universe seems to remain
transparent to light all the way until it approaches the big crunch
(thereby preventing the electromagnetic field from becoming
thermalized in between). Electromagnetic radiation into ``empty
space" would then have to be inhibited by the existence of a ``visible
dark future sky" (a time-reversed Olbers paradox), that is, by a
reduced emission power of antennae which are pointed into empty
space. This was found {\em not} to be the case \cite{33}.
The argument is not completely convincing, though, because of the
defocussing effects that must affect retarded waves through the whole
unknown lifetime of our Universe. Any ``conspirative" correlations
which would be required for their focussing onto reversed sources in the
contraction era will hardly be locally detectable now. The consistency
problem of opposite arrows in one universe may appear more severe
with respect to gravitation because of the irreversible formation
of black holes, but it is in both examples based on the unrealistic
{\em classical} field equations and does not take into account the essential
``quantum scattering" of the whole Universe at its turning point.
Since the Universe must become informationally opaque due to these
quantum effects, its ``initial" conditions (at small $a$)
for all {\em relevant} (information-carrying) degrees of freedom
appear practically free and therefore admit a condition of low entropy
at both ends of a quasitrajectory.

 We emphasize that the presumed time dependence of the density matrix
  used in (1) may completely describe an arrow of time, regardless of
   any interpretation in terms of probabilistic ``quantum events"
 (or their time ordering). The occurrence of ``events" is in fact described
in the form of {\em decoherence} by means of the smooth
Schr\"odinger dynamics \cite{34}, while the (very general \cite{1})
Zwanzig type coarse graining $\hat{P}$ used in the definition of the
entropy functional (1) has simply to be chosen compatible with it.
A monotonic increase of the corresponding entropy is then described
by a ``master equation" of the form
\begin{equation} \left(\frac{d(\hat{P}\rho)}{dt}\right)\approx
      \left(\frac{d(\hat{P}\rho)}{dt}\right)_{master}
      := \frac{\hat{P}e^{-i\hat{L}\Delta t}\hat{P}\rho -
      \hat{P}\rho}{\Delta t}\approx -i\hat{P}\hat{L}\hat{P}\rho
      -\hat{G}_{ret}\hat{P}\rho, \label{8} \end{equation}
with the Liouville operator $\hat{L}:=[H,\ldots]$,
 a positive time scale $\Delta t$ which is larger than some
  ``relaxation time" \cite{35,36}, and a positive operator
  $\hat{G}_{ret}$. It can be derived as an approximation from the
 Schr\"odinger (von Neumann) equation for $\rho$ by assuming an
 appropriate initial condition $\rho_i=\rho(t_i)$ (precisely as in the
 derivation of Boltzmann's equation from Newton's). Using
 square roots in the diagonal form of
  $\hat{G}_{ret}$, this master equation can also be
 written as a Lindblad equation.

       If the master equation holds for a coarse graining of the specific form
\begin{equation} \hat{P}\rho:= \sum_{n} P_{\alpha_n}\rho P_{\alpha_n},
 \label{9} \end{equation}
where the $P_{\alpha_n}$ form a complete set of orthogonal projection operators
 on the Hilbert space of the quantum states, the familiar
  expression
\begin{equation} p_{\alpha_1 \ldots \alpha_n} =\mbox{Tr}
    \left(C_{\alpha_n\ldots \alpha_1}\rho_i(C_{\alpha_n
    \ldots\alpha_1})^{\dagger}\right), \label{10} \end{equation}
for probabilities of time-ordered series of ``events"
 $\alpha_1(t_1), \alpha_2(t_2) \ldots\alpha_n(t_n)$, with
\begin{equation} C_{\alpha_n\ldots\alpha_1}:= e^{-iH(t_f-t_n)}
    P_{\alpha_n}e^{-iH(t_n-t_{n-1})} \ldots P_{\alpha_2}
     e^{-iH(t_2-t_1)}P_{\alpha_1}e^{-iH(t_1-t_i)}, \label{11}
 \end{equation}
defines probabilities for ``consistent histories"
 in the sense of Griffiths \cite{37}. Relaxation times correspond to the
  (extremely short) decoherence times in this case, while the projectors
  $P_{\alpha_i}$ may in general be moderately time-dependent.
They may be chosen to project onto the stable pointer basis
for the coarse-grained degrees of freedom \cite{36}.
   Such consistent histories result from a successive application
of Fermi's probabilistic Golden Rule (as used in steps of $\Delta t$
in (8)), which simply neglects interference of the probability
amplitudes after the occurrence of assumed ``quantum events."
This formal ``consistency" of {\em probabilities for histories}
must not, however, mislead to circumventing the quantum measurement
problem by interpreting the Feynman path integral as representing
an {\em ensemble} of paths or histories from which an element
can be ``picked out" by a mere increase of knowledge. This would be
as mistaken as simply replacing the wave function of an electron
by the corresponding probability distribution of particle positions.

If the master equation (8) holds for $\hat{P}\rho$, while
$\rho$ itself obeys the von Neumann equation
$id\rho/dt=\hat{L}\rho$ with an
    appropriate initial condition, the probabilities (10)
 are not changed by inserting in addition the final density matrix
$\rho_f=e^{-iH(t_f-t_i)}\rho_ie^{iH(t_f-t_i)}$
    in order to obtain a symmetric {\em form},
\begin{equation} p_{\alpha_1\ldots \alpha_n} =
\left(\mbox{Tr}
    \left[\rho_f C_{\alpha_n\ldots\alpha_1}\rho_i
    (C_{\alpha_n\ldots\alpha_1})^{\dagger}\right]\right)^{1/2}.
   \label{12} \end{equation}
This can be seen as follows. Using (\ref{9}), (\ref{10}), the
cyclic property of the trace, and $P_{\alpha_k}P_{\alpha_l}
=\delta_{kl}P_{\alpha_k}$, one has
\begin{eqnarray*}
& & \mbox{Tr}
     \left(\rho_f C_{\alpha_n\ldots\alpha_1}\rho_i
    (C_{\alpha_n\ldots\alpha_1})^{\dagger}\right) \\
  & = &
  \mbox{Tr} \left(e^{-iH(t_n-t_i)}\rho_ie^{iH(t_n-t_i)}
  P_{\alpha_n}\ldots\rho_i\ldots P_{\alpha_n}\right)\\
  & = & \mbox{Tr}\left( P_{\alpha_n}\rho(t_n)P_{\alpha_n}
  P_{\alpha_n} \ldots \rho_i\ldots P_{\alpha_n}\right)\\
  & = & \mbox{Tr}\left(\hat{P}\rho(t_n)P_{\alpha_n}\ldots
  \rho_i\ldots P_{\alpha_n}\right).
  \end{eqnarray*}
Taking into account the ``successive probabilities"
$p_{\alpha_n}:= \mbox{Tr}(P_{\alpha_n}\rho(t_n)) =p_{\alpha_1\ldots
\alpha_n}$ which arise from the master dynamics, this
expression is equal to
    $(p_{\alpha_1\ldots \alpha_n})^2$, in accordance with
    (12).
  The symmetric form
   is thus based on the factual asymmetry
which is represented by the master equation. It is caused by the special
{\em initial} condition. The probabilities would in general be changed
drastically, however, by inserting $\rho_f$ in this or a similar
symmetric expression \cite{37,38,39} if the master equation
did {\em not} hold as an approximation through all times from
$t_i$ to $t_f$. It would, in particular, not hold if the
entropy (1) with respect to a fixed relevance concept
$\hat{P}$ were low both at $t_i$ and $t_f$, thus forming a
thermodynamically time-symmetric (though possibly unitarily evolving)
universe (cf. also Page \cite{40}).

In the Heisenberg picture (and without any collapse)
$\rho_i$ and $\rho_f$ are identical, while the introduction of two
{\em independent} density matrices
$\rho_i$ and $\rho_f$ in order to define a symmetric
 ``transition probability"
 $\mbox{Trace}(\rho_f\rho_i)$ (with or without considering
``histories"), would interpret the whole Universe as {\em one}
probabilistic ``scattering event." This is, of course,
particularly dubious
in the absence of external observers.

\section{Consequences for black holes}

The dominating aspect characterizing the low entropy of the initial Universe
 seems to be its homogeneity. This has been expressed by means of the Weyl
 tensor hypothesis \cite{5} which excludes inhomogeneous past
  singularities from spacetime.
   In a thermodynamically time-symmetric
  Universe, as derived from quantum gravity, such a condition
would then also have to apply to formal ``future" singularities
which ultimately are constrained by the boundary condition.
We emphasize that the singularity theorems of classical
general relativity do not apply here since
in quantum gravity there is no classical spacetime
which could obey the Einstein equations.

How this could be achieved even in the presence of massive spherical
 black holes (for which Hawking radiation can be neglected) has
  been described elsewhere \cite{1,3}. Their external Schwarzschild metric
is static (invariant under translations of the Schwarzschild time $t$).
Because of the diverging gravitational time dilation of collapsing
matter, the space-like hypersurfaces characterized by this time
coordinate will approach the turning point of the cosmic expansion
at spatial infinity before any collapsing spherical dust shell
(or the surface of a collapsing star) has reached the horizon
that is expected to form. Classically it would then very soon
(as measured in proper time) have to pass it, and to collapse
further into a singularity, thus demonstrating the incompatibility
of exactly spherical black holes with a thermodynamically
time symmetric {\em classical} universe.
Spherical black holes are, however, compatible with a time symmetric
{\em quantum} universe.

Less symmetric matter concentrations could in principle,
even according to the classical theory, enter a
  time-symmetric and ``informationally opaque" state at extremely
   high density, from which they would have to ``grow hair" again
by means of the advanced radiation that must become relevant in the
collapse era of the Universe. Since such a state of matter appears
classically not very realistic, genuine quantum effects
(``inconsistent histories") appear indispensable close to the
Schwarzschild time which corresponds to the turning point of a
time symmetric closed universe. Although hard to interpret in a
classical picture, they are readily described by reflected
(scattered) wave packets which (1) solve the Wheeler-DeWitt equation
and (2) are compatible with the wave mechanical low entropy (``initial")
boundary condition at $a\to 0$ \cite{20}.

We emphasize that quantum cosmology with a boundary condition
of low entropy for $a\to 0$ can immediately solve many of the
problems of the classical
gravitational theory. The first concerns the
``information loss paradox" for black holes, which does not
occur because of the
absence of horizons. Hawking radiation would always stay in a
pure, although highly correlated, quantum state.
 Furthermore, since no singularities
form (except for the cosmological one) the principle of
cosmic censorship is automatically implemented. Finally,
a time-symmetric quantum Universe would prevent the occurrence of
 mass inflation inside a rotating black hole, since no Cauchy horizon could
  ever form. The cosmological scenario from mass inflation \cite{41}
   would then become obsolete.

The continuation of the classical concept of time beyond the turning
 point can thus only be formal. If the ``psychological arrow of time"
  is determined by the thermodynamical one, the Universe can only be
   observed expanding. In particular, ``information-gaining
   systems" (observers) cannot {\em continue} to exist from the expansion
into the collapse era. The different quasiclassical branches of the
wave function which are connected by ``quantum scattering" at the turning
point should rather be interpreted as all representing
different {\em expanding} universes, which disappear at the
turning point by means of
 destructive interference (similar to their coming into
existence as separate Everett branches from a symmetric
initial state at the big bang).

\vspace{3mm}
\begin{center}
{\bf Acknowledgement}
\end{center}
We wish to thank Heinz-Dieter Conradi for many helpful discussions.
We also thank David Hind for helpful suggestions and a critical
reading of the manuscript.

\end{document}